\documentclass[letterpaper,10pt,twocolumn]{article}
\usepackage{parskip}
\usepackage[font={small}]{caption}
\usepackage[hidelinks]{hyperref}
\usepackage[margin=2.5cm]{geometry}
\usepackage{amsmath,amssymb,amsfonts,amsthm,enumerate,multirow}
\usepackage[affil-it]{authblk}
\usepackage{listings}
\usepackage{graphicx} 
\usepackage{amsmath} 
\usepackage{amssymb}  
\usepackage{color}
\usepackage{caption}
\usepackage{subcaption}
\usepackage{siunitx}
\usepackage{hyperref}
\usepackage{floatrow}
\usepackage{etoolbox}
\usepackage{subcaption}
\usepackage{esint} 

\usepackage{algorithm,algorithmic,url,listings}
\newcommand*{\doi}[1]{\href{https://doi.org/#1}{doi: #1}} 

\date{}

\title{Implementation of Traction Constraints in Bragg-edge Neutron Transmission Strain Tomography}
\author[1]{J.N. Hendriks\thanks{\url{Johannes.Hendriks@newcastle.edu.au}}}
\author[1]{A.W.T. Gregg}
\author[1]{C.M. Wensrich}
\author[1]{A. Wills}

\affil[1]{School of Engineering, The University of Newcastle, Callaghan NSW 2308, Australia.}

\begin{document}

\newcommand{\coverTitle}{Implementation of traction constraints in Bragg-edge neutron transmission strain tomography}
\newcommand{\coverAuthors}{J. N. Hendriks, A. Gregg, C. Wensrich, and A. Wills}
\newcommand{\coverStatus}{Published.}

\begin{titlepage}
    \begin{center}
        {\large \em Technical report}
        
        \vspace*{2.5cm}
        %
        {\Huge \bfseries \coverTitle  \\[0.4cm]}
        
        %
        {\Large \coverAuthors \\[2cm]}
        
        \renewcommand\labelitemi{\color{red}\large$\bullet$}
        \begin{itemize}
            \item {\Large \textbf{Please cite this version:}} \\[0.4cm]
            \large
            \coverAuthors. \coverTitle. \textit{Strain}, 55(5):e12325, 2019. \doi{10.1111/str.12325}
        \end{itemize}
        
        \vfill

        
        \vfill
    \end{center}
\end{titlepage}
\twocolumn[
  \begin{@twocolumnfalse}
  \maketitle
\begin{abstract}
Several recent methods for tomographic reconstruction of stress and strain fields from Bragg-edge neutron strain images have been proposed in the literature.  This paper presents an extension of a previously demonstrated approach based on Gaussian Process regression which enforces equilibrium in the method. This extension incorporates knowledge of boundary conditions, primarily boundary tractions, into the reconstruction process. This is shown to increase the rate of convergence and is more tolerant of systematic errors that may be present in experimental measurements.  An exact expression for a central calculation in this method is also provided which avoids the need for the approximation scheme that was previously used.  Convergence of this method for simulated data is compared to existing approaches and a reconstruction from experimental data is provided.  Validation of the results to conventional constant wavelength strain measurements and comparison to prior methods shows a significant improvement. 
\vspace{3mm}
\end{abstract}
\end{@twocolumnfalse}]



\section{INTRODUCTION}
\label{sec:introduction}
Tomographic reconstruction determines a map of an unknown quantity within an object from lower dimensional projections.
A well known example is Computed Tomography (CT) where a set of flat two dimensional X-ray images are analysed to build a three dimensional image of the scalar density.
{\color{black}In the area of experimental mechanics, new techniques for high resolution imaging of strain have created significant interest in associated tomographic reconstruction processes (\cite{korsunsky2011strain,korsunsky2006principle,abbey2009feasibility,abbey2012reconstruction,kirkwoodbragg,kirkwood2015neutron,woracek2011neutron,wensrich2016granular,gregg2017tomographic,gregg2018resid,hendriks2017bragg,lionheart2015diffraction,aben2008modern})}. 
In contrast to conventional CT, tomographic reconstruction of strain seeks to determine the rank-2 tensor strain field from a set of scalar two dimensional projections of the strain field.

\begin{figure}[!ht]
    \centering
    \includegraphics[width=0.65\linewidth]{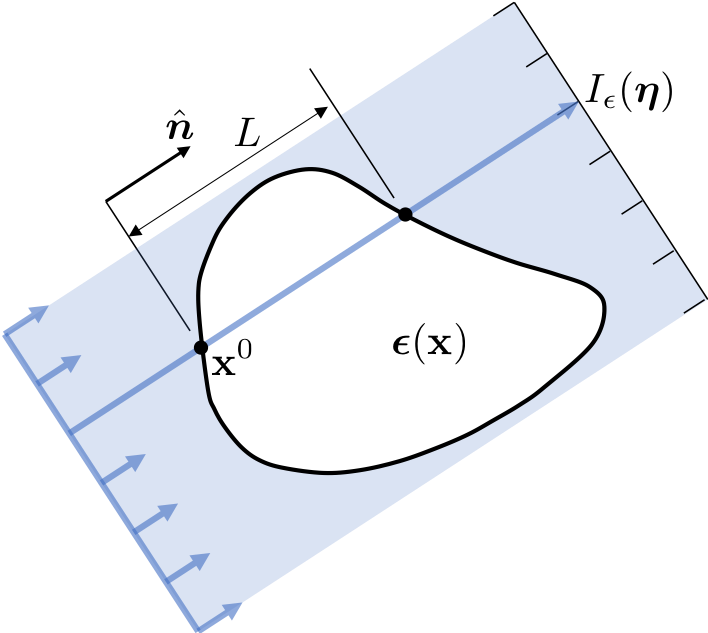}
    \caption{Coordinate system and geometry for the LRT. Each pixel of a strain image provides a measurement of the through thickness average normal strain in the direction of the ray, $I_\epsilon(\boldsymbol\eta)$. }
    \label{fig:LRT_geom}
\end{figure}

Although other approaches exist, energy-resolved neutron transmission has become a prominent method for strain imaging.
This technique analyses the relative transmission of a neutron pulse with a known wavelength-intensity spectrum through a sample over an array of detector pixels (Refer to Figure~\ref{fig:LRT_geom}).
In particular, the relative positions of sudden increases in transmission intensity as a function of wavelength---known as Bragg-edges--- can be related to strain \cite{santisteban2002strain}.

Bragg-edges are a consequence of coherent scattering; edge positions are related to lattice spacing within the sample through Bragg's law \cite{santisteban2002engineering}. 
In short, neutrons passing through a polycrystalline material can be coherently scattered by crystal planes of a certain spacing until their wavelength corresponds to a scattering angle of $180^\circ$ (i.e. backscattering). Above this wavelength no further coherent scattering occurs creating a sudden increase in relative transmission. 
Multiple Bragg-edges can be found in the transmission spectra corresponding to various lattice spacings within the samples crystal structure \cite{santisteban2002strain,santisteban2002engineering}.  
In particular, the relative shift in associated wavelength can be used to measure strain of the form;
\begin{equation}
    \langle \epsilon \rangle = \frac{\lambda - \lambda_0}{\lambda_0}
\end{equation}
where $\langle \epsilon \rangle$ is the normal strain in the transmission direction averaged over the irradiated volume, $\lambda$ is the measured wavelength of the Bragg-edge, and $\lambda_0$ is the same wavelength in an unstressed sample. As with all diffraction techniques, strain measured in this manner refers only to the elastic component.

The relationship between strain measured at each pixel and the strain field within the sample can be modelled by the Longitudinal Ray Transform (LRT) \cite{lionheart2015diffraction};
\begin{equation}\label{eq:LRT}
    I_\epsilon(\boldsymbol{\eta}) = \frac{1}{L}\int\limits^L_0 \hat{\boldsymbol{n}}^T\boldsymbol{\epsilon}(\mathbf{x}^0 + \hat{\boldsymbol{n}}s)\hat{\boldsymbol{n}}\,\mathrm{d}s,
\end{equation}
where $\boldsymbol\eta = \{\hat{\boldsymbol{n}},\mathbf{x}^0\}$. 
This maps the rank-2 tensor strain field $\boldsymbol\epsilon$ to the average normal component of strain, $I_\epsilon \in \mathbb{R}$, along the ray with direction $\hat{\boldsymbol{n}} \in \mathbb{R}^3$, entering the sample at $\mathbf{x}^0 \in \mathbb{R}^3$ and with irradiation length $L \in \mathbb{R}$, see Figure~\ref{fig:LRT_geom}. 

Inversion of the LRT is the heart of the strain tomographic reconstruction problem.

The LRT is a non-injective mapping \cite{lionheart2015diffraction}; strain fields producing a given set of strain images are not unique.
Therefore, additional information is required in order to isolate the correct (physical) strain field from all possibilities.
Previous work has considered the inclusion of compatibility and equilibrium constraints (e.g. see \cite{wensrich2016bragg} and \cite{jidling2018probabilistic}).

Compatibility was assumed in prior work focussing on special cases \cite{abbey2009feasibility,abbey2012reconstruction,wensrich2016granular,wensrich2016bragg,hendriks2017bragg}.
Axisymmetric systems were explored in \cite{abbey2009feasibility,abbey2012reconstruction}, where various basis functions were used to reconstruct strain within quenched cylinders and standard VAMAS ring-and-plug system.
Outside of axisymetric systems, \cite{wensrich2016granular} assumed compatibility in a granular system to uniquely relate the average strain within each granule to the measurements.
In \cite{wensrich2016bragg} an algorithm was developed for the reconstruction of a planar strain problem subject to \textit{in-situ} loading and then demonstrated on experimental data in \cite{hendriks2017bragg}. 
Here, the assumption of compatibility was used to uniquely relate the boundary deformations to the internal strain field, allowing for the reconstruction of the exclusively elastic strain field.

While compatibility cannot always be assumed, equilibrium must always be satisfied in physical systems. 
Equilibrium was initially used to develop two separate algorithms for axisymmetric systems \cite{kirkwood2015neutron,gregg2017tomographic}. 
In \cite{kirkwood2015neutron} equilibrium was imposed through boundary conditions. 
In contrast, \cite{gregg2017tomographic} satisfied equilibrium by minimising strain energy.

The restriction to strain fields that satisfy equilibrium fully resolves the issue of uniqueness. 
After Sharafutdinov \cite{sharafutdinov1994integral}, the null space of the LRT is known to consist of the set of strain fields that are gradients of continuous displacements which vanish at the boundary.
The well known uniqueness of solutions to linear elastic boundary value problems (due to Kirchoff \cite{knops2012uniqueness}) would then imply the only member of this set also satisfying equilibrium is the trivial one.

More general methods suitable for reconstructing planar residual strain fields have been presented in \cite{gregg2018resid} and \cite{jidling2018probabilistic}---both cases enforce equilibrium. 
In \cite{gregg2018resid} a finite series of basis functions were fit to the measurements and equilibrium was enforced at a grid of points through constraints placed on the function derivatives.
Whereas, in \cite{jidling2018probabilistic} the strain field is reconstructed with a probabilistic non-parametric approach---Gaussian Process (GP) regression---where equilibrium is enforced over the entire field by relating the strains to Airy's stress functions, and an appropriate approximation method was used to avoid numerical evaluation of a double line integral.

Conceptually both \cite{gregg2018resid} and \cite{jidling2018probabilistic} reconstruct the strain field by fitting functions. 
The differences lie in how this is achieved; a GP based method has the potential for greater expressiveness as it is not limited to a finite set of basis functions. 
However, this advantage is somewhat reduced by the use of an approximation method.

In this paper, we extend the method presented in \cite{jidling2018probabilistic} to incorporate knowledge of the tractions at the samples boundaries, this is demonstrated not only to improve the reconstruction near the boundary but over the entire field. 
Further, we provide an alternate solution to the problem of numerical integration which avoids the need for the approximation method.
A convergence comparison of the methods given in \cite{wensrich2016bragg}, \cite{gregg2018resid}, \cite{jidling2018probabilistic}, and the extended method is presented for the reconstruction of the Saint-Venant approximate cantilevered beam strain field \cite{beer2010mechanics}, and finally a comparison of applying \cite{gregg2018resid}, \cite{jidling2018probabilistic}, and the extended method to reconstruct a residual strain field from experimental data is provided.

\section{RECONSTRUCTION ALGORITHM} 
\label{sec:method}
The focus of this work is on the reconstruction of a strain field given a set of Bragg-edge strain images where each pixel represents a measurement of the form \eqref{eq:LRT}.
We extend the GP regression method presented in \cite{jidling2018probabilistic} which enforces equilibrium to resolve the uniqueness issue.
This extension utilises knowledge of boundary tractions to improve the accuracy of reconstruction.
Salient points of the method are summarised as:
\begin{enumerate}
    \item GP regression provides a non-parametric method for fitting functions to data and a revision is provided in Section~\ref{sec:gaussian_processes}.
    \item A suitable GP prior is selected for the strain field ensuring that only functions satisfying equilibrium are contained in the distribution (Section~\ref{sec:GP_strainfield}).
    \item The joint distribution between the measurements and the strain field is determined (Section~\ref{sec:measurement_joint_distribution}).
    \item The joint distribution is extended to include traction constraints (Section~\ref{sec:traction_constraints}).
    \item Estimates of the strain field are given by conditioning the GP prior on the known tractions and measurements.
\end{enumerate}

\subsection{GAUSSIAN PROCESS REGRESSION} 
\label{sec:gaussian_processes}
{\color{black}
GP regression is a Bayesian approach to linear inverse problems under the assumptions of Gaussian noise with a Gaussian prior \cite{rasmussen2006gaussian,robert2014machine,nasrabadi2007pattern}.
Although many Bayesian and non-Bayesian approaches exist to inverse problems (e.g. \cite{tarantola2005inverse,vogel2002computational}), we have adopted the GP approach to this problem for the following reasons:
\begin{enumerate}
    \item Flexibility: It employs an infinite dimensional basis expansion to represent the unknown function being estimated, thus it accommodates a highly flexible class of estimators. At the same time, this infinite dimensional expansion is never explicity used since it is only necessary to evaluate the spatial correlation of the function at the user defined points of interest. This avoids the problem of having to choose an optimal finite set of basis functions.
    \item Non-parametric estimator: GP regression provides an estimate of the unknown function that relies on the spatial variables and the stochastic variables.
    Once the user specifies the spatial locations, then an estimate at these locations can be computed using standard finite-dimensional conditional Guassian machinery. This allows estimates of a strain field to be computed that are not sensitive to the specific choice of resolution or points of interest. 
    \item Natural inclusion of linear constraints: GPs are closed under affine transformation (see e.g. \cite{jidling2017linearly}), and therefore allow linear constraints, such as equilibrium and line-integral observations, to be easily accommodated. 
\end{enumerate}
For an in-depth discussion on GPs see \cite{rasmussen2006gaussian}, however a short summary detailing the basic elements is provided below. }

A GP is a generalisation of the multivariate Gaussian probability distribution and to a Gaussian distribution over functions. 
This distribution is uniquely defined over the spatial coordinate $\mathbf{x}$ by a mean function $\mathbf{m}(\mathbf{x}) = \mathbb{E}\left[\mathbf{f}(\mathbf{x})\right]$ and a covariance function $\mathbf{K}(\mathbf{x},\mathbf{x}') = \mathbb{E}\left[\left(\mathbf{f}(\mathbf{x})-\mathbf{m}(\mathbf{x})\right)\left(\mathbf{f}(\mathbf{x}')-\mathbf{m}(\mathbf{x}')\right)^T\right]$, where $\mathbb{E}$ refers to the expected value.

{\color{black}The covariance function $\mathbf{K}(\mathbf{x},\mathbf{x}')$ encodes our prior belief in the properties of the underlying function, such as its smoothness. 
In general, any function that generates a positive definite symmetric covariance can be used.
In this work we exclusively use the squared-exponential $K(\mathbf{x},\mathbf{x}') = \sigma_f^2\exp(-\frac{1}{2}|\mathbf{x}-\mathbf{x}'|^2/l^2)$,} which corresponds to a Bayesian linear regression model with an infinite number of basis functions \cite{rasmussen2006gaussian}. {\color{black}Note that the prior variance, $\sigma_f$, and the length-scale, $l$ are not chosen manually, but rather optimised for by maximising the log-likelihood of the data (see Chapter~5 in \cite{rasmussen2006gaussian}). This adapts the imposed smoothness to best fit the data.}

GP regression provides a predictor function that can be used to estimate function values at specified locations, $\mathbf{x}_*$, based on a set of measurements that can be expressed as linear functionals of the form, $y = \mathcal{V}_\mathbf{x}[\mathbf{f}(\mathbf{x})] + e$, and $e$ is zero-mean Gaussian noise of standard deviation $\sigma_n$.
In this case, any finite set of measurements, $\mathbf{Y} = \{y_1,\dots,y_N\}$, for inputs $\mathbf{x}_1,\dots,\mathbf{x}_N$ and the function value evaluated at $\mathbf{x}_*$ are jointly Gaussian;

\small
\begin{equation}
    \begin{bmatrix}
        y_1 \\ \vdots \\ y_N \\ \mathbf{f}(\mathbf{x}_*)
    \end{bmatrix}\sim \mathcal{N}\left(\begin{bmatrix}\boldsymbol\mu_y \\ \mathbf{m}(\mathbf{x}_*)\end{bmatrix},
    \begin{bmatrix}
             \mathbf{K}_{\mathbf{y}\mathbf{y}'}+\sigma_n^2 I & \mathbf{K}_{\mathbf{y}*} \\
           \mathbf{K}_{\mathbf{y}*}^T & \mathbf{K}(\mathbf{x}_*,\mathbf{x}_*)
    \end{bmatrix}
    \right),
\end{equation} \normalsize
where 
\small
\begin{equation}
\begin{split}
    \boldsymbol\mu =  \begin{bmatrix}
        \mathcal{V}_{\mathbf{x}_1}\mathbf{m}(\mathbf{x}_1) \\ \vdots \\ \mathcal{V}_{\mathbf{x}_N}\mathbf{m}(\mathbf{x}_N) 
        \end{bmatrix}, \quad 
    \mathbf{K}_{\mathbf{y}*}(\mathbf{x}_*) = \begin{bmatrix}
        \mathcal{V}_{\mathbf{x}_1}\mathbf{K}(\mathbf{x}_1,\mathbf{x}_*) \\
        \vdots \\
        \mathcal{V}_{\mathbf{x}_N}\mathbf{K}(\mathbf{x}_N,\mathbf{x}_*)
    \end{bmatrix}
\end{split}
\end{equation}\normalsize

and
\small
\begin{equation}
\begingroup 
\setlength\arraycolsep{1pt}
    \mathbf{K}_{\mathbf{y}\mathbf{y}'} = \begin{bmatrix}
        \mathcal{V}_{\mathbf{x}_1}\mathbf{K}(\mathbf{x}_1,\mathbf{x}_1)\mathcal{V}_{\mathbf{x}_1}^T & \cdots & \mathcal{V}_{\mathbf{x}_1}\mathbf{K}(\mathbf{x}_1,\mathbf{x}_N)\mathcal{V}_{\mathbf{x}_N}^T  \\
         \vdots & \ddots & \vdots \\ 
         \mathcal{V}_{\mathbf{x}_N}\mathbf{K}(\mathbf{x}_N,\mathbf{x}_1)\mathcal{V}_{\mathbf{x}_1}^T & \cdots & \mathcal{V}_{\mathbf{x}_N}\mathbf{K}(\mathbf{x}_N,\mathbf{x}_N)\mathcal{V}_{\mathbf{x}_N}^T \\ 
    \end{bmatrix}
\endgroup
\end{equation}\normalsize
An estimate, $\boldsymbol{\mu}_{\mathbf{f}_*|\mathbf{Y}}(\mathbf{x}_*)$ of $\mathbf{f}(\mathbf{x}_*)$ and its variance, $\Sigma_{\mathbf{f}_*|\mathbf{Y}}(\mathbf{x}_*)$ based on the measurements can then be given by;

\small
\begin{equation}\label{eq:GPR}
    \begin{split}
        \boldsymbol{\mu}_{\mathbf{f}_*|\mathbf{Y}}(\mathbf{x}_*) &= \mathbf{m}(\mathbf{x}_*) + \mathbf{K}_{\mathbf{y}*}(\mathbf{x}_*)^T\\ &\hspace{15mm}\left(\mathbf{K}_{\mathbf{yy}'}+\sigma_n^2 I\right)^{-1}(\mathbf{Y}-\boldsymbol\mu_y), \\
        \Sigma_{\mathbf{f}_*|\mathbf{Y}}(\mathbf{x}_*) &= \mathbf{K}(\mathbf{x}_*,\mathbf{x}_*)- \mathbf{K}_{\mathbf{y}*}(\mathbf{x}_*)^T\\ &\hspace{15mm}\left(\mathbf{K}_{\mathbf{yy}'}+\sigma_n^2 I\right)^{-1}\mathbf{K}_{\mathbf{y}*}(\mathbf{x}_*).
    \end{split}
\end{equation}\normalsize

{\color{black}
The benefits of employing the GP approach are not in computing the numerical solution;  as discussed in \cite{rasmussen2006gaussian}, equation~\ref{eq:GPR} coincides with a variety of other methods (e.g. Tikhonov regularization).  Rather the benefit lies in the approach to the formulation of the problem to begin with.
This formulation does not require the problem to rendered finite dimensional while still providing a solution with feasible computational complexity. 
A finite dimensional formulation can be achieved in at least two ways, but they are accompanied by the following issues:
\begin{itemize}
\item Use of a finite basis function expansion: this has the disadvantage of requiring the user to choose the most appropriate bases for the current problem, which may not be obvious.
\item Spatial discretisation of the strain field: this again has the disadvantage of requiring another user choice for the resolution and pattern of the discretisation. In addition, this approach does not easily afford the inclusion of equilibrium constraints, for example.
\end{itemize}}


\subsection{STRAIN FIELD GAUSSIAN PROCESS} 
\label{sec:GP_strainfield}
In this section we design a GP for two-dimensional strain fields with mean and covariance function defined in a way that ensures the distribution only spans functions satisfying equilibrium.  This is done by following the procedure outlined in \cite{jidling2018probabilistic}. 

We define a GP for a distribution of Airy's Stress functions, $\varphi(\mathbf{x})$, from which we can define stress as $\sigma_{xx}(\mathbf{x}) = \frac{\partial^2\varphi(\mathbf{x})}{\partial y^2}$, $\sigma_{yy}(\mathbf{x}) = \frac{\partial^2\varphi(\mathbf{x})}{\partial x^2}$, and $\sigma_{xy}(\mathbf{x}) = \frac{\partial^2\varphi(\mathbf{x})}{\partial x \partial y}$. 
From Hooke's law, under an assumption of plane stress, this provides strain of;
\begin{equation}\label{eq:phitoeps}
    \boldsymbol{\mathcal{E}}(\mathbf{x}) =  \underbrace{\begin{bmatrix}\frac{\partial^2}{\partial y^2}-\nu\frac{\partial^2}{\partial x^2}\\ -(1+\nu)\frac{\partial^2}{\partial x \partial y} \\ \frac{\partial^2}{\partial x^2}-\nu\frac{\partial^2}{\partial y^2} \end{bmatrix}}_{\mathcal{L}_\mathbf{x}} \psi(\mathbf{x}) = \mathcal{L}_{\mathbf{x}} \varphi(\mathbf{x}),
\end{equation}
where $\boldsymbol{\mathcal{E}} = [\epsilon_{xx}, \epsilon_{xy}, \epsilon_{yy}]^T$, and $\nu$ is poisson's ratio. 

The a priori distribution of $\varphi(\mathbf{x})$ is assumed to have mean $m_\varphi(\mathbf{x})=0$, covariance $K_\varphi(\mathbf{x},\mathbf{x}')$, and $\mathbf{x} = [x,y]^T$. This allows us to write a GP for the strain field with mean function $\mathbf{m}_\epsilon = \mathcal{L}_{\mathbf{x}}m_\varphi(\mathbf{x})=\mathbf{0}$ and covariance function 
\begin{equation}
    \boldsymbol{K}_{\epsilon\epsilon}(\mathbf{x},\mathbf{x}') = \mathcal{L}_{\mathbf{x}} 
    K_\varphi(\mathbf{x},\mathbf{x}')
    \mathcal{L}_{\mathbf{x}} ^T.
\end{equation}
This is possible as GPs are closed under linear functional transformations \cite{papoulis2002probability,hennig2013quasi,wahlstrom2015modeling}.

Although many options exist for the choice of covariance function $K_\varphi(\mathbf{x},\mathbf{x}')$, the squared-exponential and Matern covariance functions were both shown to be suitable in \cite{jidling2018probabilistic}.
It is also possible to build a covariance function from the finite two-dimensional Fourier basis used in \cite{gregg2018resid}, however we would then be required to specify the number of basis functions used, limiting the expressiveness of the Gaussian process. 

The shorthand $K_\varphi = K_\varphi(\mathbf{x}_i,\mathbf{x}_j)$ and $\mathbf{K}_{\epsilon\epsilon} = \mathbf{K}_{\epsilon\epsilon}(\mathbf{x}_i,\mathbf{x}_j)$ will be used where appropriate.

\subsection{MEASUREMENT JOINT DISTRIBUTION} 
\label{sec:measurement_joint_distribution}
In order to estimate the values of the strain field from a set of strain images, we require the joint distribution of the measurements, $\mathbf{I}_\epsilon$, and strains;
\begin{equation}
    \begin{bmatrix}
        \mathbf{I}_\epsilon \\ \boldsymbol{\mathcal{E}}
    \end{bmatrix} \sim 
    \mathcal{N}\left(\begin{bmatrix}
        \mathbf{0}\\ \mathbf{0}
    \end{bmatrix},\begin{bmatrix}
        \boldsymbol{K}_{II} + \sigma_n^2 I & \boldsymbol{K}_{I\epsilon} \\
         \boldsymbol{K}_{I\epsilon}^T & \boldsymbol{K}_{\epsilon\epsilon}
    \end{bmatrix}\right)
\end{equation}

The cross covariance between an LRT measurement $I_{\epsilon i}$ and strain at the $j^\text{th}$ input location $\mathbf{x}_j$ is
\begin{equation}\label{eq:Kip_intrac}
\begin{split}
    \boldsymbol{K}_{I\epsilon}(\boldsymbol\eta_i,\mathbf{x}_j) &= \frac{1}{L_i}\int\limits_0^{L_i} \bar{\boldsymbol{n}}_i\boldsymbol{K}_{\epsilon\epsilon}(\mathbf{x}_i^0 + \hat{\boldsymbol{n}}_i s,\mathbf{x}_j)\,\mathrm{d}s
\end{split}
\end{equation}
where $\bar{\boldsymbol{n}} = \begin{bmatrix} \hat{n}_1^2 & 2\hat{n}_1\hat{n}_2 & \hat{n}_2^2 \end{bmatrix}$. The covariance between two measurements, $I_{\epsilon i}$ and $I_{\epsilon j}$ is given by 
\begin{equation}
\begin{split}
    \boldsymbol{K}_{II}(\boldsymbol\eta_i,\boldsymbol\eta_j) = \frac{1}{L_iL_j}\int\limits_0^{L_i}\hspace{-1.5mm}\int\limits_0^{L_j} \bar{\boldsymbol{n}}_i\boldsymbol{K}_{\epsilon\epsilon}(&\mathbf{x}_i^0 + \hat{\boldsymbol{n}}_i s,\\
    &
    \mathbf{x}_j^0 + \hat{\boldsymbol{n}}_i s)\bar{\boldsymbol{n}}_j^T\,\mathrm{d}s_i\,\mathrm{d}s_j .
\end{split}
\end{equation}

In its current form, an analytical solution to \eqref{eq:Kip_intrac} appears intractable and \cite{jidling2018probabilistic} proposed either the use of numerical integration or an approximation scheme.
However, in the special cases where the $x$ or $y$ direction of the strain is aligned with the line integral an analytical solution is apparent (i.e. $\mathrm{d}s = \mathrm{d}x$). 
This can be used to provide a general solution by noting that strain can be rotated from a coordinate system aligned with the ray to the global coordinate system according to $\boldsymbol{\epsilon} = \mathbf{R}(\hat{\boldsymbol{n}}_i)\boldsymbol\epsilon^{\{i\}}\mathbf{R}(\hat{\boldsymbol{n}}_i)^T$,
where $\boldsymbol\epsilon^{\{i\}}$ is defined as strain expressed in the coordinates of line $i$ and
\begin{equation}
    \mathbf{R}(\hat{\boldsymbol{n}}_i) = \begin{bmatrix}
        \hat{n}_1 & -\hat{n}_2 \\ \hat{n}_2 & \hat{n}_1
    \end{bmatrix}.
\end{equation}
In vector form this can be written as
\begin{equation}
    \boldsymbol{\mathcal{E}} = \underbrace{\begin{bmatrix}
        \hat{n}_1^2 & -2\hat{n}_1\hat{n}_2 & \hat{n}_2^2 \\
        \hat{n}_2\hat{n}_{1} & \hat{n}_1^2-\hat{n}_2^2 & -\hat{n}_1\hat{n}_2 \\
        \hat{n}_2^2 & 2\hat{n}_1\hat{n}_2 & \hat{n}_1^2
    \end{bmatrix}}_{\mathbf{J}(\hat{\boldsymbol{n}}_i)}\boldsymbol{\mathcal{E}}^{\{i\}}
\end{equation}
Therefore, the covariance can now be written as
\begin{equation}\label{eq:Kip}
\begin{split}
    &\boldsymbol{K}_{I\epsilon}(\boldsymbol\eta_i,\mathbf{x}_j) = \frac{1}{L_i}\bar{\boldsymbol{n}}^{\{l\}}
    \mathcal{V}_{\mathbf{x}_i}
    K_\varphi
    \mathcal{L}_{\mathbf{x}_j} ^T \mathbf{J}(\hat{\boldsymbol{n}}_i)^T, \\
    &\mathcal{V}_{\mathbf{x}_i} = \begin{bmatrix}\int\limits_0^{L_i}\frac{\partial^2}{\partial y_i^2}\,\mathrm{d}x_i-\nu\frac{\partial}{\partial x_i}\\ -(1+\nu)\frac{\partial}{\partial y_i} \\ \frac{\partial}{\partial x_i}-\nu\int\limits_0^{L_i}\frac{\partial^2}{\partial y_i^2}\,\mathrm{d}x_i \end{bmatrix}.
\end{split}
\end{equation}
Here, $\bar{\boldsymbol{n}}^{\{l\}} = [1, 0, 0]$. The solutions to these equations are given in Appendix~\ref{sec:line_integral}.

Unfortunately, the same technique used to solve the first integral cannot be applied to the second integral as the strain cannot be simultaneously expressed in two different coordinate systems.
However, using \eqref{eq:Kip} we are required to only numerically evaluate a single integral---improving computation time and accuracy.

Extension to non-convex geometry is straightforward; the measurement model becomes a sum of integrals over each segment. Details can be found in \cite{jidling2018probabilistic}.

\subsection{BOUNDARY TRACTIONS} 
\label{sec:traction_constraints}
So far, we have a joint distribution that allows the estimation of strains satisfying equilibrium from Bragg-edge neutron strain measurements. 
The estimation of these strains can be improved by including knowledge about tractions acting on the surface.
Strain within a physical body is subject to conditions imposed upon it by the tractions on its surface. 
Firstly, tractions can be directly related to stress at the surface and hence to strain using
\begin{equation}
    T_i = C_{ijkl}\epsilon_{kl}\zeta_j,
\end{equation}
where $\mathbf{C}$ is the stiffness tensor from Hooke's law that relates strain to stress, and $\boldsymbol\zeta$ is the surface normal. 

Secondly, knowledge of the boundary tractions can inform us about the average stress (hence strain) within a solid.
The mean stress theorem \cite{cristescu2003mechanics} states that when equilibrium is satisfied, and in the absence of body forces, the average stress within a body, $\mathcal{B}$, is related to the surface integral of the tractions through
\begin{equation}
\resizebox{0.95\linewidth}{!}{$
    \frac{1}{V}\int\limits_\mathcal{B}\boldsymbol{\sigma}(\mathbf{x})\,\mathrm{d}v = \frac{1}{2V}\int\limits_{\partial\mathcal{B}}(\mathbf{x}\otimes\mathbf{T}(\mathbf{x})+\mathbf{T}(\mathbf{x})\otimes\mathbf{x})\,\mathrm{d}a.
$}
\end{equation}

While we are unlikely to know the distribution of surface tractions at a point of loading, surface areas not subject to contact loads will have zero-tractions. 
This is particularly applicable to residual strain problems where there are often no surface tractions at all. 
This knowledge is incorporated into the reconstruction process by the inclusion of artificial observations of zero traction ($\mathbf{T} = 0)$.
Assuming plane stress, these observations are related to our strain function by
\begin{equation}
    \mathbf{T} = \underbrace{\begin{bmatrix}
        \zeta_1 & \zeta_2 & 0 \\ 0 & \zeta_1 & \zeta_2
    \end{bmatrix}}_{
    \tilde{\boldsymbol{\zeta}}}\underbrace{\frac{E}{1-\nu^2}
    \begin{bmatrix}
        1 & 0 & \nu \\
        0 & 1-\nu & 0 \\
        \nu & 0 & 1
    \end{bmatrix}}_{\mathbf{C}_{2D}}\boldsymbol{\mathcal{E}}.
\end{equation}

The covariance between traction observations, the cross covariance between traction observations and strains, and the cross covariance between LRT measurements and traction observations are given by
\begin{equation}
\begin{split}
    &\boldsymbol{K}_{TT}(\mathbf{x}_i,\mathbf{x}_j) = \tilde{\boldsymbol\zeta}\mathbf{C}_{2D}\boldsymbol{K}_{\epsilon\epsilon}(\mathbf{x}_i,\mathbf{x}_j)\mathbf{C}_{2D}^T\tilde{\boldsymbol\zeta}^T\\
    &\boldsymbol{K}_{T\epsilon}(\mathbf{x}_i,\mathbf{x}_j) = \tilde{\boldsymbol\zeta}\mathbf{C}_{2D}\boldsymbol{K}_{\epsilon\epsilon}(\mathbf{x}_i,\mathbf{x}_j) \\
    & \boldsymbol{K}_{IT} = \boldsymbol{K}_{I\epsilon}(\boldsymbol\eta_i,\mathbf{x}_j) \mathbf{C}_{2D}^T\tilde{\boldsymbol\zeta}^T
\end{split}
\end{equation}

The joint distribution of the strains, LRT measurements, and the zero-traction observations is then given by
\begin{equation}
    \begin{bmatrix}
        \mathbf{I}_\epsilon \\ \mathbf{T} \\ \boldsymbol{\mathcal{E}}
    \end{bmatrix} \sim 
    \mathcal{N}\left(\begin{bmatrix}
        \mathbf{0}\\ \mathbf{0} \\ \mathbf{0}
    \end{bmatrix},\begin{bmatrix}
        \boldsymbol{K}_{II}+\sigma_n^2 \mathbf{I} & \boldsymbol{K}_{IT} & \boldsymbol{K}_{I\epsilon} \\ \boldsymbol{K}_{IT}^T & \boldsymbol{K}_{TT}+\sigma_t^2\mathbf{I} & \boldsymbol{K}_{T\epsilon} \\ \boldsymbol{K}_{I\epsilon}^T & \boldsymbol{K}_{T\epsilon}^T & \boldsymbol{K}_{\epsilon\epsilon}
    \end{bmatrix}\right)
\end{equation}
{\color{black}where $\sigma_t$ is the uncertainty of the artificial strain measurements. Theoretically $\sigma_t=0$, however in practice it was found using a small non-zero value yielded better results particularly if the samples boundary was not known perfectly. The optimal value of $\sigma_t$ can be found along with the other parameters ($\sigma_f$ and $l$) by maximising the marginal log likelihood.} Strain estimates can now be made using Equation~\eqref{eq:GPR} incorporating both measurements from strain images and knowledge of the boundary tractions.


\section{DEMONSTRATION---SIMULATION } 
\label{sec:simulation_results}

Reconstructing the strain field of the classical cantilevered beam has been examined in \cite{wensrich2016bragg,jidling2018probabilistic} and \cite{gregg2018resid}.
As such, we use this example to demonstrate the improvement that incorporating boundary tractions has on the reconstruction.
A direct comparison of the convergence of these methods as well as a Gaussian process method that includes boundary tractions is provided.

\begin{figure}[!ht]
    \centering
    \includegraphics[width=0.8\linewidth]{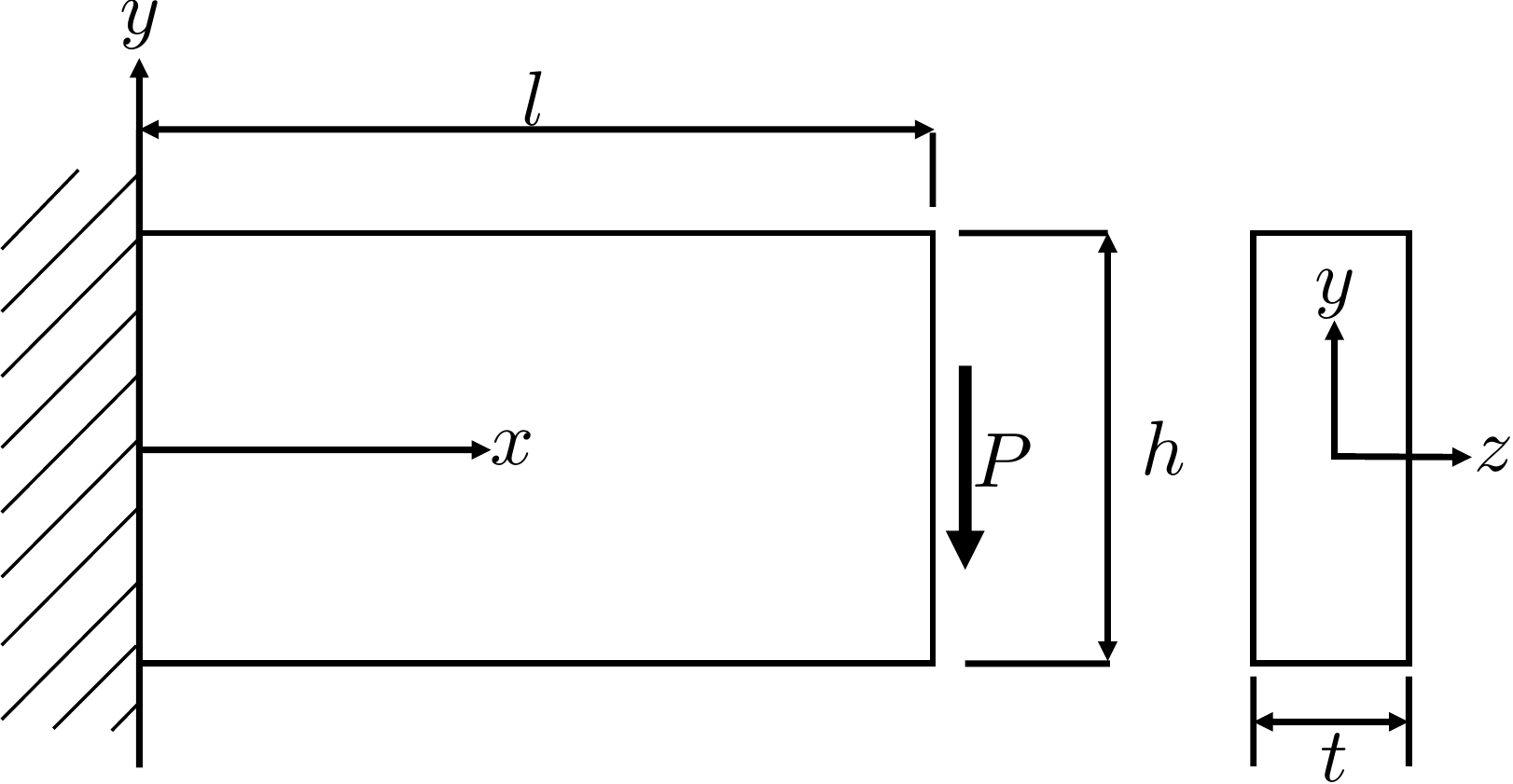}
    \caption{Cantilever beam geometry and coordinate system with $l = 20\text{mm}$, $h=10\text{mm}$, $t=5\text{mm}$, $E = 200\text{Gpa}$, $P = 2\text{KN}$, $\nu = 0.3$, and $I = \frac{th^3}{12}$.}
    \label{fig:CB_geom}
\end{figure}

The classical cantilevered beam is shown in Figure~\ref{fig:CB_geom} and the Saint-Venant approximation \cite{beer2010mechanics} to the strain field {\color{black}under the assumption of plane stress} is:
\begin{equation}
    \boldsymbol{\mathcal{E}}(\mathbf{x}) = \begin{bmatrix}
        \frac{P}{EI}(L-x)y  \\
        -\frac{(1+\nu)P}{2EI}\left(\left(\frac{h}{2}\right)^2-y^2\right) \\
        -\frac{\nu P}{EI}(L-x)y
    \end{bmatrix}
\end{equation}

Strain images were numerically simulated from this field assuming 512 pixels per projection and a detector width of $28\text{mm}$ (corresponding to Micro-Channel Plate (MCP) detector specifications currently used in experiments \cite{tremsin2012high}).
The values at each pixel correspond to a measurement of the form \eqref{eq:LRT} and were corrupted with zero-mean Gaussian noise of standard deviation $\sigma = 1.25\times10^{-4}$.

\begin{figure*}[!ht]
    \centering
    \includegraphics[width=0.65\linewidth]{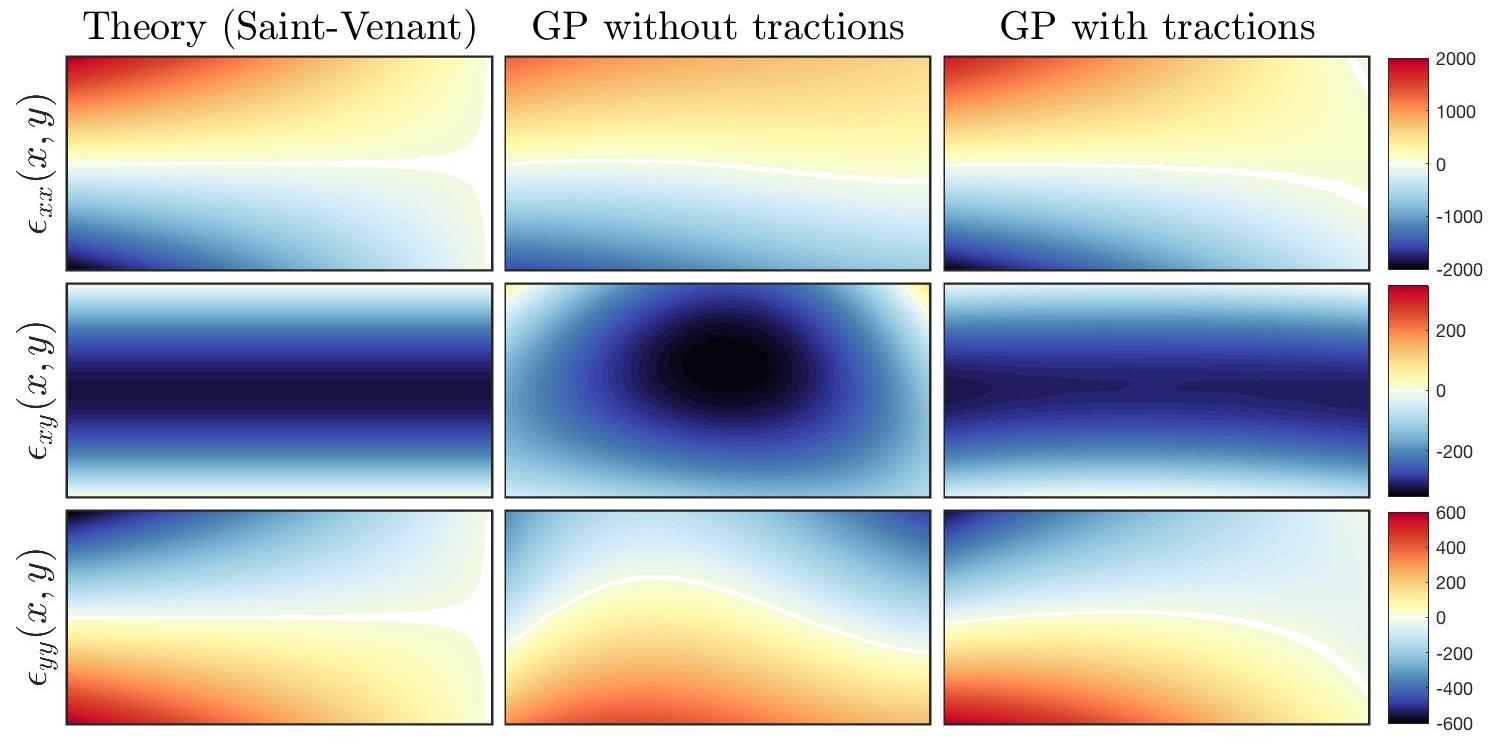}
    \caption{(Left) The Saint-Venant strain field from which measurements were generated. (Center) Reconstruction from a limited measurement set using the Gaussian process method presented in \cite{jidling2018probabilistic}. (Right) Reconstruction from a limited measurement set using the Gaussian process method extended to include boundary tractions. Value in $\mu$Strain.}
    \label{fig:CB_tract_demo}
\end{figure*}

To demonstrate the benefit of including boundary tractions, reconstruction of the strain field from a limited set of measurements was performed with and without the inclusion of tractions (Figure~\ref{fig:CB_tract_demo}). 
The subset of measurements consists of 40 pixels from each of 4 different projections.
Without the inclusion of boundary tractions the reconstruction shows significant error, particularly the $\epsilon_{xy}$ component.
Zero-traction measurements were then included at a total of 100 points along the top and bottom surface of the cantilevered beam (where it is known to have zero-traction), and the subsequent reconstruction shows considerable improvement.

\begin{figure}[!h]
    \centering
    \includegraphics[width=0.85\linewidth]{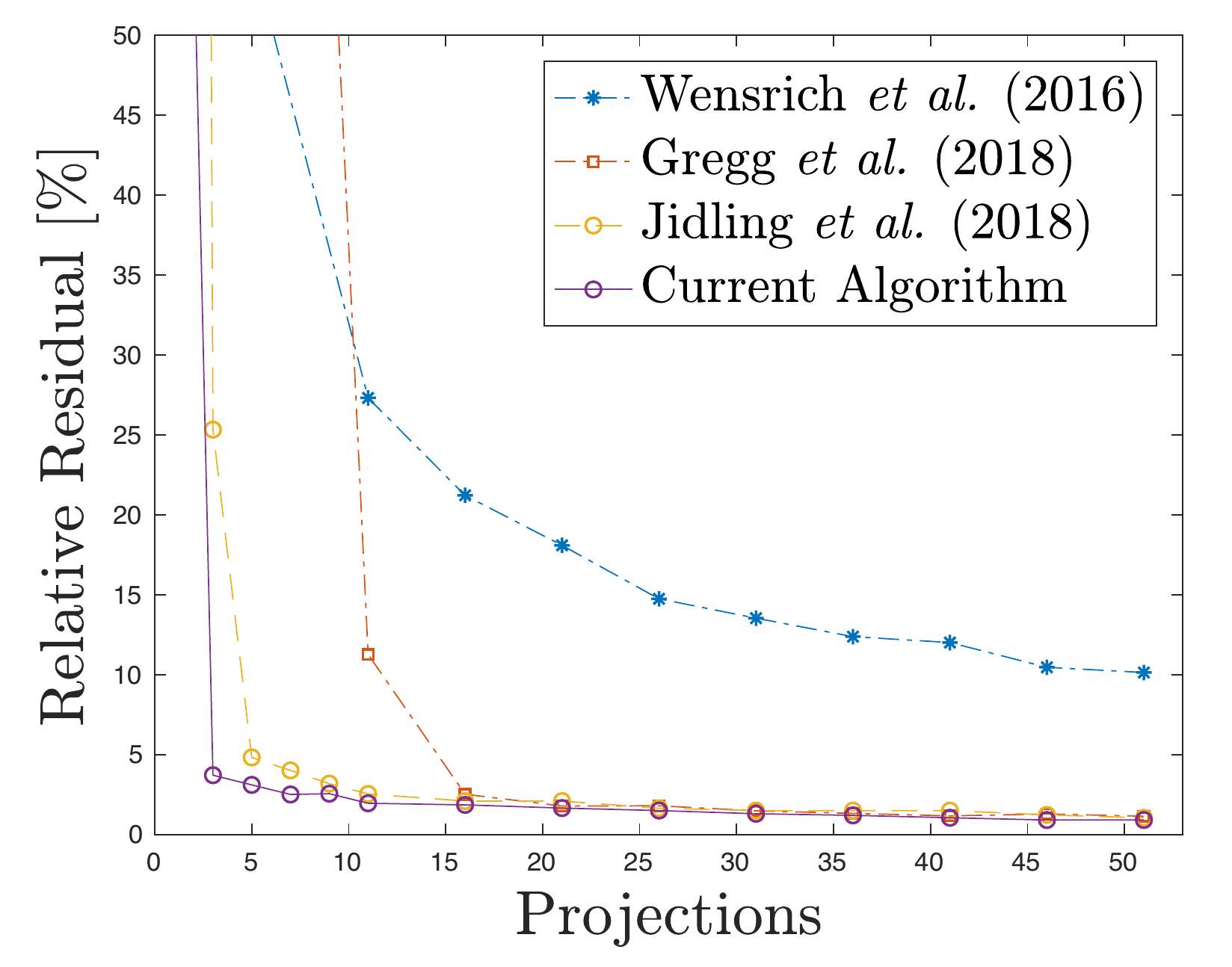}
    \caption{The convergence of the current algorithm compared to the boundary reconstruction method \cite{wensrich2016bragg}, the finite basis method \cite{gregg2018resid}, and the Gaussian process regression method demonstrated in \cite{jidling2018probabilistic}.}
    \label{fig:convergence}
\end{figure}

Convergence of the algorithm is shown in Figure~\ref{fig:convergence} and compared to the boundary reconstruction method \cite{wensrich2016bragg}, a finite basis method \cite{gregg2018resid} with $n=6$ wave numbers, and a Gaussian process method that does not include traction measurements \cite{jidling2018probabilistic}. 
For each reconstruction, projection angles were evenly spaced over $180^\circ$ degrees.
Both Gaussian process regression methods show significantly faster convergence than the other algorithms, with the method including tractions converging fastest with less than $5\%$ relative error for three projections---the minimum number of unique projections angles required in two dimensions \cite{wensrich2016bragg}.
For 51 projections the methods excluding the boundary reconstruction approach show good convergence to the theoretical solution. 
With the current algorithm giving a marginally better relative error of 0.9\% compared to $1.05\%$ and $1.15\%$ for the methods from \cite{jidling2018probabilistic} and \cite{gregg2018resid}, respectively. 

\begin{figure}[!h]
    \centering
    \includegraphics[width=0.85\linewidth]{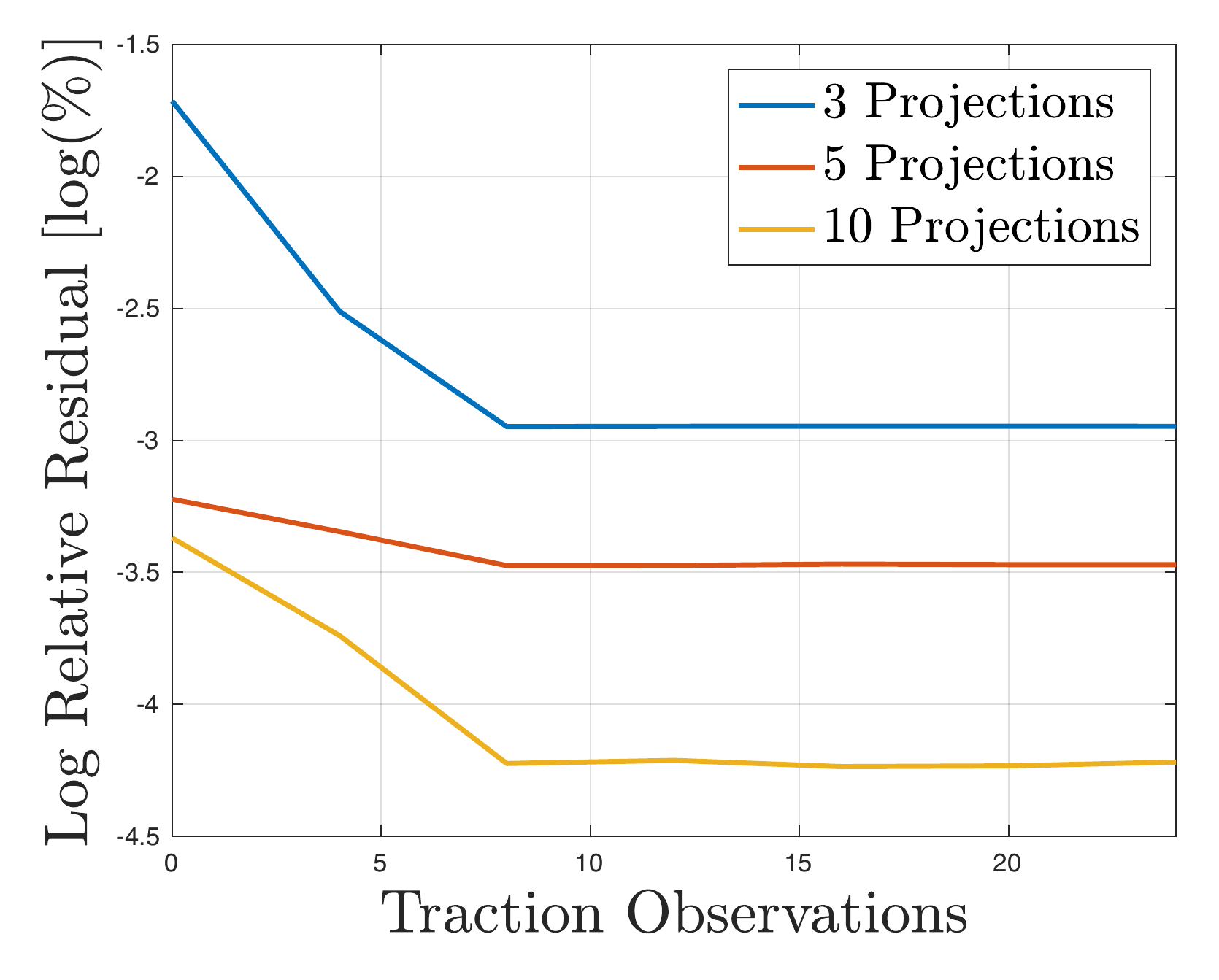}
    \caption{Log relative residual of the current algorithm as the total number of traction observations is increased. Results are shown for reconstructions made from three, five, and ten projections. }
    \label{fig:trac_study}
\end{figure}

{\color{black}The decrease in the relative residual with respect to the total number of traction observations is shown in Figure~\ref{fig:trac_study}. 
Results are shown for three measurements sets taken from the convergence study (Figure~\ref{fig:convergence}) corresponding to three, five, and 10 projections. 
The results indicate the rate of decrease is independent of the number of projections and that for this case no further improvement is achieved after 8 traction observations.
In general, the number of traction points required depends on the length scale which is not known till after the algorithm is run.
Therefore it is suggested to include a fine spacing of traction observations in locations where the boundary tractions are known to be zero.}

\section{EXPERIMENTAL DEMONSTRATION } 
\label{sec:experimental_results}
\begin{figure}[!ht]
    \centering
    \includegraphics[width=0.4\linewidth]{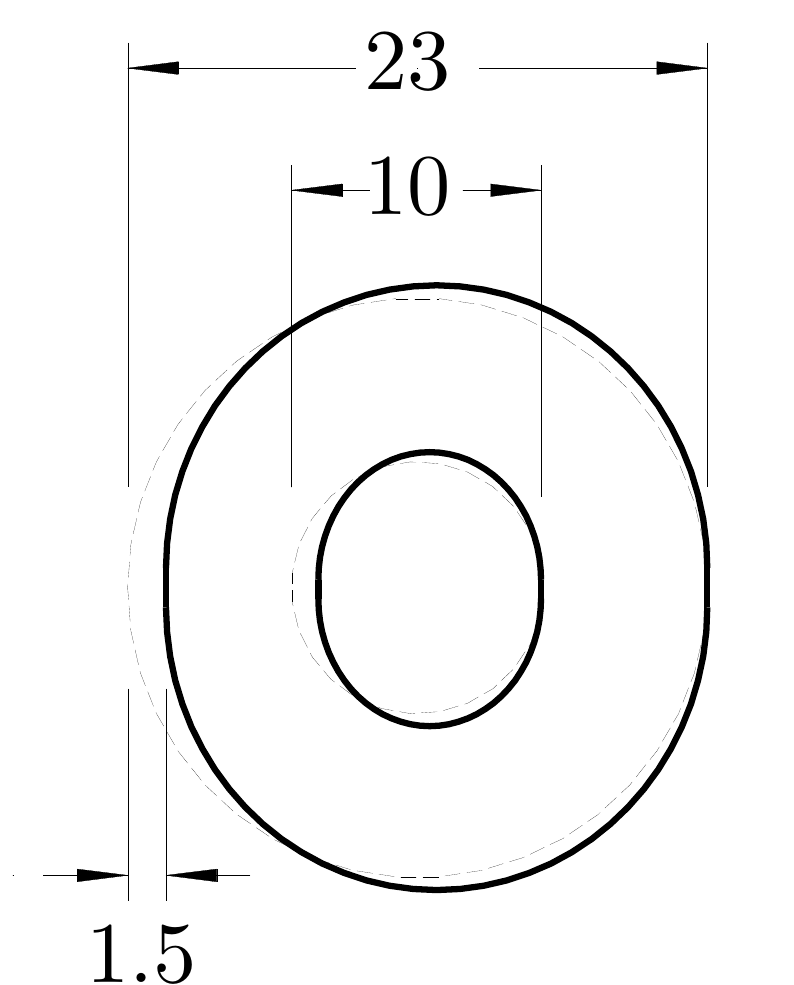}
    \caption{Crushed ring sample geometry. Dimensions are in mm.}
    \label{fig:sample_geometry}
\end{figure}

The presented tomographic reconstruction method was then applied to a set of strain images measured from a plastically deformed steel ring.  This \textit{crushed ring} sample formed the focus of a strain tomography experiment carried out on RADEN (Energy Resolved Neutron Imaging) instrument at the Japan Proton Accelerator Research Complex (J-PARC) \cite{shinohara2015commissioning,nakajima2017materials} in January 2018.
{\color{black}All details of the sample design, experiment set-up, and strain measurement can be found in \cite{gregg2018resid}.  For context, we provide the following summary;

The crushed ring sample was formed by first heat treating an EN26 steel cylinder ($23\text{mm}$ outer diameter, $10\text{mm}$ inner diameter and 14\text{mm} thick).  This heat treatment removed all prior residual stress and resulted in a final hardness of $290\text{HV}$. 
After heat treatment, the hollow cylinder was plastically deformed by $1.5\text{mm}$ (see Figure~\ref{fig:sample_geometry}) using approximately $8.4\text{KN}$ of load.
The load was subsequently removed leaving an approximately two-dimensional, (i.e. plane stress) residual strain field within the sample.

A set of 50 strain images were measured over golden angle increments of rotation using an MCP detector ($512\times 512$ pixels, $55\text{$\mu$m}$ per pixel) at a distance of $17.9\text{m}$ from the source. 
Measurements were performed over columns of pixels spanning the thickness of the sample to provide one-dimensional projections of the two-dimensional strain field.
Strain measurements were made as an average over each column using the method described in \cite{santisteban2002strain}. 
In total, 20,664 measurements of the form \eqref{eq:LRT} were collected over the 50 projections, each with a measurement uncertainty around $\sigma_n = 1\times 10^{-4}$.}

\begin{figure*}[!ht]
    \centering
    \includegraphics[width=0.8\linewidth]{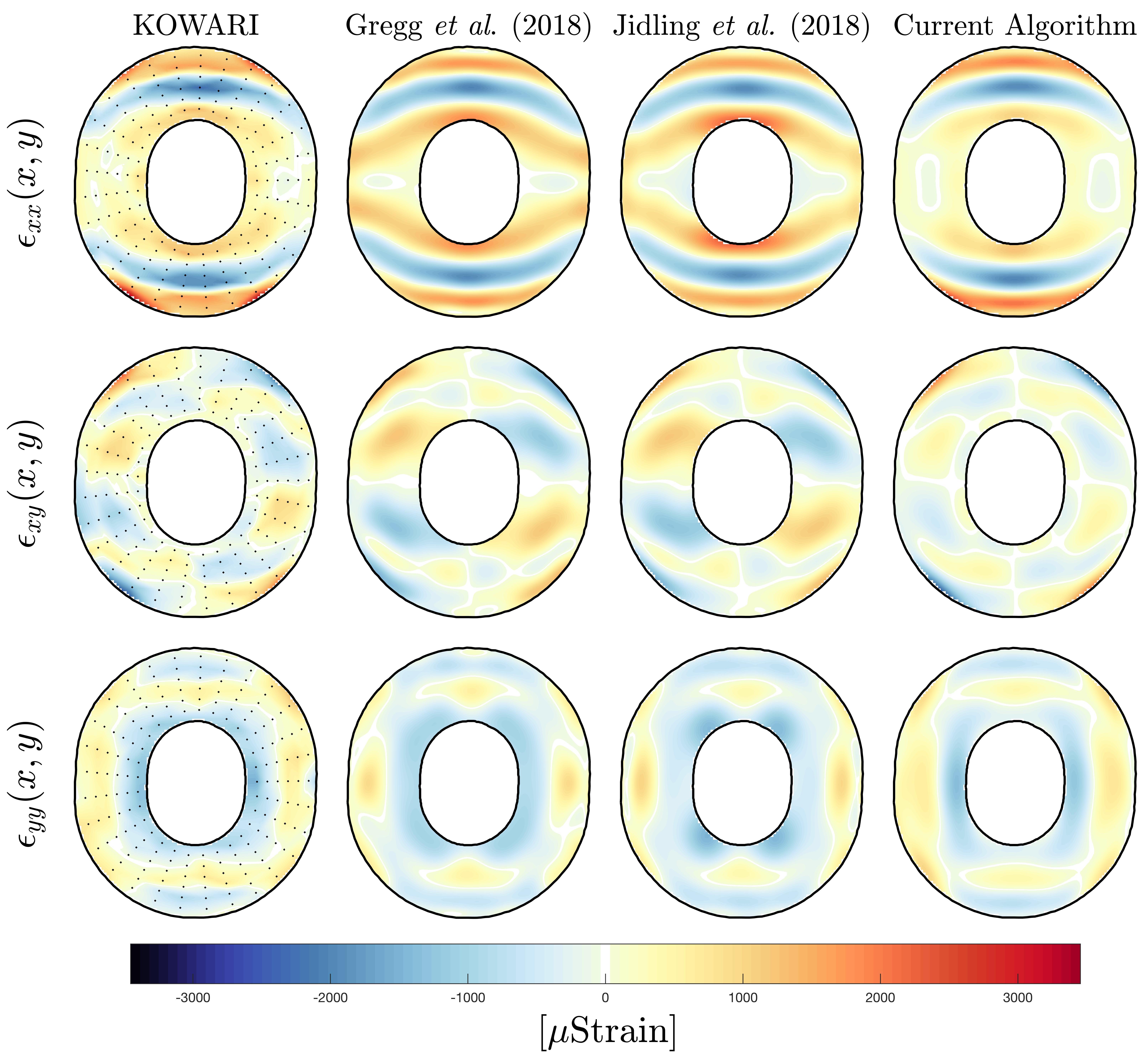}
    \caption{(Left) Strain fields interpolated from point-wise KOWARI strain scans (markers indicate measurement locations) compared to reconstructions from strain images taken using RADEN. (Center left) reconstruction presented in Gregg \textit{et al.} (2018) \cite{gregg2018resid}. (Center right) Reconstruction using the method presented in Jidling \textit{et al.} (2018)\cite{jidling2018probabilistic}. (Right) Current Algorithm that includes boundary tractions.}
    \label{fig:crushed_ring_results}
\end{figure*}

The strain field reconstructed from these measurements is shown in Figure~\ref{fig:crushed_ring_results} alongside prior reconstructions from the same data by Gregg \textit{et al.} \cite{gregg2018resid}, and Jidling \textit{et al.} \cite{jidling2018probabilistic}.  Each of these prior reconstructions were based on an equilibrium constraint alone.

{\color{black}The reconstruction from the current algorithm was based on 200 zero-traction measurements evenly distributed over both the interior and exterior boundaries.}

Also shown is a detailed conventional strain scan (e.g. \cite{kisi2012applications,fitzpatrick2003analysis,noyan2013residual}) measured using the KOWARI constant wavelength strain-diffractometer at the Australian Nuclear Science and Technology Organisation (ANSTO) \cite{kirstein2009strain,brule2006residual,kirstein2010kowari}.
{\color{black}This strain-scan provide measurements of the three in-plane components of strain over a mesh of points within the sample using a $0.5\times 0.5 \times 14\text{mm}$ gauge volume. 
Data points shown in the figure indicate the measurement locations.
The uncertainty of each measurement in this scan (estimated from the peak fitting process) was in the order of $\sigma = 7.5\times 10^{-5}$. }
All details of this measurement can be found in \cite{gregg2018resid}.

Comparison of the current reconstruction to the KOWARI strain scan shows close agreement; qualitatively better than the prior methods \cite{jidling2018probabilistic,gregg2018resid}.
In particular, the banding in the $\epsilon_{xx}$ component has been reduced in-line with the KOWARI results, and concentrations in the $\epsilon_{yy}$ component have been reduced with the square shaped tensile region present in the KOWARI results now being captured in the reconstruction.

{\color{black}
Figure~\ref{fig:resid_error} provides the $\epsilon_{xx}$ discrepancy between the three methods of reconstruction and the KOWARI strain fields. 
When interpreting the results in Figure~\ref{fig:crushed_ring_results} and Figure~\ref{fig:resid_error}, it should be noted that the KOWARI strain fields are interpolated (and extrapolated on the boundaries) from a grid of point-wise measurements and these interpolations are not bound by equilibrium or the zero boundary traction condition.
Hence, the KOWARI strain fields should not be treated as absolute ground truth but rather as a reasonable point of comparison.
\begin{figure*}[!ht]
    \centering
    \includegraphics[width=0.7\linewidth]{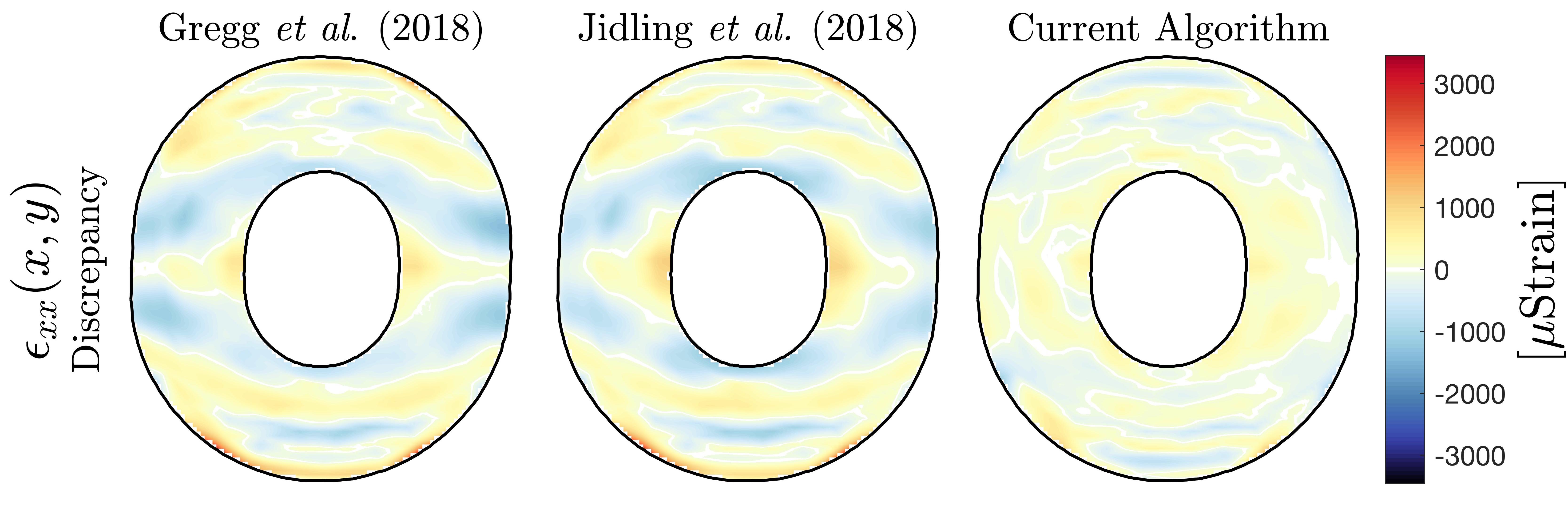}
    \caption{\color{black}Discrepancy in the between the strain field interpolated from point-wise KOWARI measurements and the three reconstruction techniques. Only the $\epsilon_{xx}$ component is shown; the $\epsilon_{xy}$ and $\epsilon_{yy}$ components show similar trends.}
    \label{fig:resid_error}
\end{figure*}

The expected magnitudes of the discrepancy between the three reconstruction methods and the KOWARI strain measurements were calculated as follows (each given in terms of the $\epsilon_{xx}$, $\epsilon_{xy}$, and $\epsilon_{yy}$ components respectively);
\begin{itemize}
    \item{Gregg \textit{et al.} \cite{gregg2018resid}; $333\mu\epsilon$, $309\mu\epsilon$, and $239\mu\epsilon$.}  
    \item{Jidling \textit{et al.} \cite{jidling2018probabilistic}; $339\mu\epsilon$, $296.9\mu\epsilon$, and $282\mu\epsilon$.}
    \item{Current algorithm; $213\mu\epsilon$, $206\mu\epsilon$, and $178\mu\epsilon$.}
\end{itemize}}
Although reduced by the addition of traction constraints, some error still exists. 

{\color{black}It should also be noted that the discrepancies highlighted above are well in excess of the posterior estimate of uncertainty from equation \eqref{eq:GPR} (in the order of $3\times 10^{-5}$). 
This estimate of variance assumes that the only source of measurement error is random additive Gaussian noise.
This further supports the suggestion made in \cite{gregg2018resid} that there exists systematic errors in the Bragg-edge strain measurements that are contributing to the reconstruction error. }
Potential sources of these errors are discussed in detail in \cite{gregg2018resid}, and are summarised as: a thickness-dependent shift in edge-centres, possibly due to beam hardening \cite{vogel2000rietveld}, resulting in psuedo-strains present in the measurements; complications in the edge fitting process due to strain gradients; and the combined effects of beam divergence and potential sample misalignment.
Clearly the impact of these systematic errors on the reconstruction has been reduced by including knowledge of the boundary tractions into the algorithm.
Characterisation and removal of these systematic errors is an area of continued research.

{\color{black}
It must also be said that the assumption of a two-dimensional plane stress state is a potential source of error.
While all care was taken to ensure the stress field was dominated by the in-plane components, the geometry of the sample is such that a small amount of deviation from this ideal case is to be expected.  This deviation is likely to have some small impact on the reconstruction.
}


\section{CONCLUSION AND FUTURE WORK} 
\label{sec:conclusion}
A method for incorporating known boundary tractions into an algorithm suitable for reconstructing two-dimensional strain fields from Bragg-edge neutron strain images has been presented.
The inclusion of boundary tractions reduces the number of measurements required and improves the algorithms ability to reject noise; in particular, systematic error. 
The method we have developed ensures that the entire field satisfies equilibrium with no assumptions of strain compatibility, and is therefore suitable for reconstructing residual strain fields.

As an additional outcome, an analytical solution has been provided to the first integral required to calculate the measurement covariance in the GP method. This eliminates the need to approximate the GP by a finite set of basis functions, removing the need to choose the number of basis functions and maintaining a greater expressiveness.

This method was demonstrated in simulation, where the inclusion of boundary tractions was shown to significantly improve the result when a limited measurement set was used.
Furthermore, a convergence comparison was provided highlighting the benefit of a Gaussian process based method over other available approaches with a GP method including boundary tractions converging fastest.

Results from a reconstruction of a residual strain field within a crushed ring from experimental data collected on the RADEN energy resolved neutron imaging instrument were presented.
These results showed closer agreement with strain scans measured using the KOWARI constant wavelength diffractometer when compared with previous methods.
This reduction in error demonstrates that the inclusion of boundary tractions can reduce the impact of systematic measurement error in the reconstruction process.

Future work will include extending the Gaussian process regression method to three-dimensions. 
In three dimensions, the Beltrami stress functions \cite{sadd2009elasticity} provide a general analogy to two-dimensional Airy stress functions and can potentially perform the same function in the design of a suitable kernel function.
The method presented for including traction constraints can also be trivially extended to three-dimensions.
This is not without significant additional complexity in terms of both numerical and experimental concerns, however it does demonstrate that the path to a three dimensional extension of work is apparent.

\section{ACKNOWLEDGEMENTS}
This work is supported by the Australian Research Council through a Discovery Project Grant (DP170102324). Access to the RADEN and KOWARI instruments was made possible through the respective user access programs of J-PARC and ANSTO (J-PARC Long Term Proposal 2017L0101 and ANSTO Program Proposal PP6050). The authors would also like to thank AINSE Limited for providing financial assistance (PGRA) and support to enable work on this project.

\appendix                                     
\section{Line Integral Equations}\label{sec:line_integral}
Expanding the terms given in \eqref{eq:Kip} yields
\begin{equation*}
       \boldsymbol{K}_{I\epsilon}(\boldsymbol{\eta}_i,\mathbf{x}_j) = \frac{1}{L_i}
    \begin{bmatrix} \Gamma_{1} & \Gamma_{2} & \Gamma_{3}\end{bmatrix}
    \mathbf{J}(\hat{\boldsymbol{n}}_i)^T
\end{equation*}
Where
\begin{equation*}
    \begin{split}
        \Gamma_{1} &= \nu^2\left[\frac{\partial^3}{\partial x_i \partial x_j^2}K_\psi\right]^{L_i}_0 - 2\nu \left[\frac{\partial^3}{\partial x_i \partial y_j^2}K_\psi\right]^{L_i}_0 \\
        & \qquad \qquad \qquad \qquad \qquad \qquad  +\int\limits_0^L\frac{\partial^4}{\partial y_i^2 \partial y_j^2}K_\psi\,\mathrm{d}x_i, \\
        \Gamma_{2} &= -\nu(\nu+1)\left[\frac{\partial^3}{\partial y_i\partial x_j^2}K_\psi\right]_0^{L_i} \\
        &\qquad \qquad \qquad \qquad \qquad + (\nu+1)\left[\frac{\partial^3}{\partial y_i \partial y_j^2}K_\psi\right]_0^{L_i}, \\
        \Gamma_{3} &=  -\nu\left[\frac{\partial^3}{\partial x_i \partial x_j^2}K_\psi\right]^{L_i}_0 +(1+\nu^2) \left[\frac{\partial^3}{\partial x_i \partial y_j^2}K_\psi\right]^{L_i}_0 \\
        & \qquad \qquad \qquad \qquad \qquad \qquad - \nu\int\limits_0^{L_i}\frac{\partial^4}{\partial y_i^2 \partial y_j^2}K_\psi\,\mathrm{d}x_i . 
    \end{split}
\end{equation*}
Here, we have used the fact that $\frac{\partial}{\partial y_i \partial x_j} = \frac{\partial}{\partial x_i \partial y_j}$.

If the covariance function $K_\psi(\mathbf{x}_i,\mathbf{x}_j)$ is chosen to be the squared-exponential then;

\begin{equation*}
\begin{split}
    &\frac{\partial^3}{\partial x_i \partial x_j^2}K_\psi = \frac{3}{l_1^4}(x_i-x_j)K_\psi - \frac{1}{l_1^6}(x_i-x_j)^3K_\psi, \\
    &\frac{\partial^3}{\partial x_i \partial y_j^2}K_\psi= \frac{1}{l_1^2l_2^2}(x_i-x_j)K_\psi \\
    &\qquad \qquad \qquad \qquad - \frac{1}{l_1^2l_2^4}(x_i-x_j)(y_i-y_j)^2K_\psi, \\
    &\frac{\partial^3}{\partial y_i \partial y_j^2}K_\psi = \frac{3}{l_2^4}(y_i-y_j)K_\psi - \frac{1}{l_2^6}(y_i-y_j)^3K_\psi, \\
    &\frac{\partial^3}{\partial y_i \partial x_j^2}K_\psi = \frac{1}{l_1^2l_2^2}(y_i-y_j)K_\psi \\
    &\qquad \qquad \qquad \qquad - \frac{1}{l_1^4l_2^2}(x_i-x_j)^2(y_i-y_j)K_\psi, \\
    &\int\limits_0^{L_i}\frac{\partial^4}{\partial y_i^2 \partial y_j^2}K_\psi\,\mathrm{d}x_i = C_y\int\limits_0^{L_i} K_\psi \, \mathrm{d}x_i \\
    &C_y = \left(\frac{1}{l_2^8}(y_i-y_j)^4 - \frac{6}{l_2^6}(y_i-y_j)^2 + \frac{3}{l_2^4}\right) \\
    & \int\limits_0^{L_i} K_\psi \, \mathrm{d}x_i = \Bigg[-\sigma_f^2\sqrt{\frac{\pi}{2}}l_1 \exp\left(\frac{-(y_i-y_j)^2}{2l_2^2}\right) \\
    & \qquad \qquad \qquad \qquad \qquad \qquad \qquad \qquad \text{erf}\left(\frac{x_j-x_i}{\sqrt{2}l_1}\right)\Bigg]_0^{L_i} 
\end{split}
\end{equation*}


\bibliographystyle{ieeetr}
\bibliography{References}

\begin{thebibliography}{10}

\bibitem{korsunsky2011strain}
A.~M. Korsunsky, N.~Baimpas, X.~Song, J.~Belnoue, F.~Hofmann, B.~Abbey, M.~Xie,
  J.~Andrieux, T.~Buslaps, and T.~K. Neo, ``Strain tomography of
  polycrystalline zirconia dental prostheses by synchrotron x-ray
  diffraction,'' {\em Acta Materialia}, vol.~59, no.~6, pp.~2501--2513, 2011.

\bibitem{korsunsky2006principle}
A.~M. Korsunsky, W.~J. Vorster, S.~Y. Zhang, D.~Dini, D.~Latham, M.~Golshan,
  J.~Liu, Y.~Kyriakoglou, and M.~J. Walsh, ``The principle of strain
  reconstruction tomography: Determination of quench strain distribution from
  diffraction measurements,'' {\em Acta Materialia}, vol.~54, no.~8,
  pp.~2101--2108, 2006.

\bibitem{abbey2009feasibility}
B.~Abbey, S.~Y. Zhang, W.~J. Vorster, and A.~M. Korsunsky, ``Feasibility study
  of neutron strain tomography,'' {\em Procedia Engineering}, vol.~1, no.~1,
  pp.~185--188, 2009.

\bibitem{abbey2012reconstruction}
B.~Abbey, S.~Y. Zhang, W.~Vorster, and A.~M. Korsunsky, ``Reconstruction of
  axisymmetric strain distributions via neutron strain tomography,'' {\em
  Nuclear Instruments and Methods in Physics Research Section B: Beam
  Interactions with Materials and Atoms}, vol.~270, pp.~28--35, 2012.

\bibitem{kirkwoodbragg}
H.~J. Kirkwood, B.~Abbey, H.~M. Quiney, S.~Y. Zhang, A.~S. Tremsin, and
  A.~Korsunsky, ``Bragg edge neutron strain tomography,'' 2003.

\bibitem{kirkwood2015neutron}
H.~J. Kirkwood, S.~Y. Zhang, A.~S. Tremsin, A.~M. Korsunsky, N.~Baimpas, and
  B.~Abbey, ``Neutron strain tomography using the radon transform,'' {\em
  Materials Today: Proceedings}, vol.~2, pp.~S414--S423, 2015.

\bibitem{woracek2011neutron}
R.~Woracek, D.~Penumadu, N.~Kardjilov, A.~Hilger, M.~Strobl, R.~Wimpory,
  I.~Manke, and J.~Banhart, ``Neutron bragg-edge-imaging for strain mapping
  under in situ tensile loading,'' {\em Journal of Applied Physics}, vol.~109,
  no.~9, p.~093506, 2011.

\bibitem{wensrich2016granular}
C.~Wensrich, J.~Hendriks, and M.~Meylan, ``Bragg edge neutron transmission
  strain tomography in granular systems,'' {\em Strain}, vol.~52, no.~1,
  pp.~80--87, 2016.

\bibitem{gregg2017tomographic}
A.~Gregg, J.~Hendriks, C.~Wensrich, and M.~Meylan, ``Tomographic reconstruction
  of residual strain in axisymmetric systems from bragg-edge neutron imaging,''
  {\em Mechanics Research Communications}, 2017.

\bibitem{gregg2018resid}
A.~Gregg, J.~Hendriks, C.~Wensrich, A.~Wills, A.~Tremsin, V.~Luzin,
  T.~Shinohara, O.~Kirstein, M.~Meylan, and E.~Kisi, ``Tomographic
  reconstruction of two-dimensional residual strain fields from bragg-edge
  neutron imaging,'' {\em Physical Review Applied}, vol.~10, no.~6, p.~064034,
  2018.

\bibitem{hendriks2017bragg}
J.~Hendriks, A.~Gregg, C.~Wensrich, A.~Tremsin, T.~Shinohara, M.~Meylan,
  E.~Kisi, V.~Luzin, and O.~Kirsten, ``Bragg-edge elastic strain tomography for
  in situ systems from energy-resolved neutron transmission imaging,'' {\em
  arXiv preprint arXiv:1708.03426}, 2017.

\bibitem{lionheart2015diffraction}
W.~R. Lionheart and P.~J. Withers, ``Diffraction tomography of strain,'' {\em
  Inverse Problems}, vol.~31, no.~4, p.~045005, 2015.

\bibitem{aben2008modern}
H.~Aben, J.~Anton, and A.~Errapart, ``Modern photoelasticity for residual
  stress measurement in glass,'' {\em Strain}, vol.~44, no.~1, pp.~40--48,
  2008.

\bibitem{santisteban2002strain}
J.~Santisteban, L.~Edwards, M.~Fitzpatrick, A.~Steuwer, P.~Withers, M.~Daymond,
  M.~Johnson, N.~Rhodes, and E.~Schooneveld, ``Strain imaging by bragg edge
  neutron transmission,'' {\em Nuclear Instruments and Methods in Physics
  Research Section A: Accelerators, Spectrometers, Detectors and Associated
  Equipment}, vol.~481, no.~1, pp.~765--768, 2002.

\bibitem{santisteban2002engineering}
J.~Santisteban, L.~Edwards, M.~Fizpatrick, A.~Steuwer, and P.~Withers,
  ``Engineering applications of bragg-edge neutron transmission,'' {\em Applied
  Physics A}, vol.~74, no.~1, pp.~s1433--s1436, 2002.

\bibitem{wensrich2016bragg}
C.~Wensrich, J.~Hendriks, A.~Gregg, M.~Meylan, V.~Luzin, and A.~Tremsin,
  ``Bragg-edge neutron transmission strain tomography for in situ loadings,''
  {\em Nuclear Instruments and Methods in Physics Research Section B: Beam
  Interactions with Materials and Atoms}, vol.~383, pp.~52--58, 2016.

\bibitem{jidling2018probabilistic}
C.~Jidling, J.~Hendriks, N.~Wahlström, A.~Gregg, T.~B. Schön, C.~Wensrich,
  and A.~Wills, ``Probabilistic modelling and reconstruction of strain,'' {\em
  Nuclear Instruments and Methods in Physics Research Section B: Beam
  Interactions with Materials and Atoms}, vol.~436, pp.~141 -- 155, 2018.

\bibitem{sharafutdinov1994integral}
V.~A. Sharafutdinov, {\em Integral geometry of tensor fields}, vol.~1.
\newblock Walter de Gruyter, 1994.

\bibitem{knops2012uniqueness}
R.~J. Knops and L.~E. Payne, {\em Uniqueness theorems in linear elasticity},
  vol.~19.
\newblock Springer Science \& Business Media, 2012.

\bibitem{beer2010mechanics}
F.~Beer, E.~Johnston~Jr, J.~Dewolf, and D.~Mazurek, ``Mechanics of materials,
  sixth edit edition,'' 2010.

\bibitem{rasmussen2006gaussian}
C.~E. Rasmussen and C.~K. Williams, {\em Gaussian processes for machine
  learning}, vol.~1.
\newblock MIT press Cambridge, 2006.

\bibitem{robert2014machine}
C.~Robert, ``Machine learning, a probabilistic perspective,'' 2014.

\bibitem{nasrabadi2007pattern}
N.~M. Nasrabadi, ``Pattern recognition and machine learning,'' {\em Journal of
  electronic imaging}, vol.~16, no.~4, p.~049901, 2007.

\bibitem{tarantola2005inverse}
A.~Tarantola, {\em Inverse problem theory and methods for model parameter
  estimation}, vol.~89.
\newblock siam, 2005.

\bibitem{vogel2002computational}
C.~R. Vogel, {\em Computational methods for inverse problems}, vol.~23.
\newblock Siam, 2002.

\bibitem{jidling2017linearly}
C.~Jidling, N.~Wahlstr{\"o}m, A.~Wills, and T.~B. Sch{\"o}n, ``Linearly
  constrained gaussian processes,'' {\em arXiv preprint arXiv:1703.00787},
  2017.

\bibitem{papoulis2002probability}
A.~Papoulis and S.~U. Pillai, {\em Probability, random variables, and
  stochastic processes}.
\newblock Tata McGraw-Hill Education, 2002.

\bibitem{hennig2013quasi}
P.~Hennig and M.~Kiefel, ``Quasi-newton method: A new direction,'' {\em Journal
  of Machine Learning Research}, vol.~14, no.~Mar, pp.~843--865, 2013.

\bibitem{wahlstrom2015modeling}
N.~Wahlstr{\"o}m, {\em Modeling of Magnetic Fields and Extended Objects for
  Localization Applications}.
\newblock PhD thesis, Link{\"o}ping University Electronic Press, 2015.

\bibitem{cristescu2003mechanics}
N.~D. Cristescu, E.-M. Craciun, and E.~So{\'o}s, {\em Mechanics of elastic
  composites}.
\newblock CRC Press, 2003.

\bibitem{tremsin2012high}
A.~Tremsin, J.~McPhate, A.~Steuwer, W.~Kockelmann, A.~M~Paradowska,
  J.~Kelleher, J.~Vallerga, O.~Siegmund, and W.~Feller, ``High-resolution
  strain mapping through time-of-flight neutron transmission diffraction with a
  microchannel plate neutron counting detector,'' {\em Strain}, vol.~48, no.~4,
  pp.~296--305, 2012.

\bibitem{shinohara2015commissioning}
T.~Shinohara and T.~Kai, ``Commissioning start of energy-resolved neutron
  imaging system, raden in j-parc,'' {\em Neutron news}, vol.~26, no.~2,
  pp.~11--14, 2015.

\bibitem{nakajima2017materials}
K.~Nakajima, Y.~Kawakita, S.~Itoh, J.~Abe, K.~Aizawa, H.~Aoki, H.~Endo,
  M.~Fujita, K.~Funakoshi, W.~Gong, {\em et~al.}, ``Materials and life science
  experimental facility (mlf) at the japan proton accelerator research complex
  ii: Neutron scattering instruments,'' {\em Quantum Beam Science}, vol.~1,
  no.~3, p.~9, 2017.

\bibitem{kisi2012applications}
E.~H. Kisi and C.~J. Howard, {\em Applications of neutron powder diffraction},
  vol.~15.
\newblock Oxford University Press, 2012.

\bibitem{fitzpatrick2003analysis}
M.~E. Fitzpatrick and A.~Lodini, {\em Analysis of residual stress by
  diffraction using neutron and synchrotron radiation}.
\newblock CRC Press, 2003.

\bibitem{noyan2013residual}
I.~C. Noyan and J.~B. Cohen, {\em Residual stress: measurement by diffraction
  and interpretation}.
\newblock Springer, 2013.

\bibitem{kirstein2009strain}
O.~Kirstein, V.~Luzin, and U.~Garbe, ``The strain-scanning diffractometer
  kowari,'' {\em Neutron News}, vol.~20, no.~4, pp.~34--36, 2009.

\bibitem{brule2006residual}
A.~Brule and O.~Kirstein, ``Residual stress diffractometer kowari at the
  australian research reactor opal: status of the project,'' {\em Physica B:
  Condensed Matter}, vol.~385, pp.~1040--1042, 2006.

\bibitem{kirstein2010kowari}
O.~Kirstein, U.~Garbe, and V.~Luzin, ``Kowari-opal's new stress diffractometer
  for the engineering community: Capabilities and first results,'' in {\em
  Materials Science Forum}, vol.~652, pp.~86--91, Trans Tech Publ, 2010.

\bibitem{vogel2000rietveld}
S.~Vogel, {\em A Rietveld-approach for the analysis of neutron time-of-flight
  transmission data}.
\newblock PhD thesis, Christian-Albrechts Universit{\"a}t Kiel, 2000.

\bibitem{sadd2009elasticity}
M.~H. Sadd, {\em Elasticity: theory, applications, and numerics}.
\newblock Academic Press, 2009.

\end{thebibliography}

\end{document}